\documentclass[11pt]{article}
\usepackage{moriond,epsfig}
\newenvironment{2figures}[1]{\begin{figure}[#1] 
    \begin{tabular}{p{.55\textwidth}p{.48\textwidth}} }
 {  \end{tabular}
 \end{figure}
}

\newcommand{\ra}{\rightarrow}
\def\etm{\mathrm{E_t\!\!\!\!/}}
\newcommand{\nc}{\newcommand}
\nc{\lsim}{\mbox{\raisebox{-.6ex}{~$\stackrel{<}{\sim}$~}}}
\nc{\gsim}{\mbox{\raisebox{-.6ex}{~$\stackrel{>}{\sim}$~}}}
\nc{\esim}{\mbox{\raisebox{-.6ex}{~$\stackrel{-}{\sim}$~}}}

\begin{document}
\vspace*{1cm}
\title{SEARCH FOR LIGHT HIGGS BOSON AT LHC \\
VIA PRODUCTION THROUGH WEAK BOSON FUSION}
\author{K.MAZUMDAR\\
 On behalf of CMS Collaboration, CERN}

\address{Experimental High Energy Physics Group, Tata Institute of Fundamental Research, Mumbai 400 005, India \& EP Division, CERN, Geneva.}
\maketitle
\abstracts{The LHC potential for observing a light Higgs boson
produced through Weak Boson Fusion mode, ${\rm qq}\rightarrow {\rm qqH}$, is presented. For non-hadronic decays modes of the Higgs boson the process is identified with a final
state containing two energetic forward-backward jets, separated with a large rapidity and a 
hadronically quiet  central region. The use of these properties, combined
with special features of some of the decay modes enhances the
potential of an early discovery of a light Higgs boson both in the Standard Model and beyond. 
The recent studies done in the context of CMS experiment are discussed.
} 
\section{Higgs Boson Discovery at LHC}
The primary goal of the large hadron collider (LHC), is to unravel the mystery of the electroweak symmetry 
breaking. The search for the Standard Model (SM) Higgs boson has therefore been one of the main 
motivations behind enormous 
efforts put in by the omni-purpose p-p experiments, ATLAS and CMS,
in the  LHC. The first 
collisions at the LHC, at a centre-of-mass energy of 14 TeV, are expected 
by June 2007. By 
2010 the accumulated luminosity will be 30 fb$^{-1}$/experiment, increased to 100 fb$^{-1}$/year eventually when the machine will run with the design luminosity of $10^{34}$ cm$^{-2}$ s$^{-1}$. 
As shown in Fig.~\ref{fig:hobs1}, the SM Higgs boson will be detected in more than 
one channel over the whole mass range ($m_{\rm H}$) from 80 GeV/c$^2$ to $\sim 1$ TeV/c$^2$. 
The experimentally favoured domain (at 95\% C.L.) is bounded from below at 114.4 GeV/c$^2$ by
direct searches at LEP and from above at  219 GeV/c$^2$ by electroweak precision measurements at LEP and SLD.
The luminosity required by CMS to cover the lower part of this favoured region  is displayed in Fig.~\ref{fig:hobs2}.
The weak boson fusion (WBF) mode for the Higgs boson production at the LHC, ${\rm qq}\rightarrow {\rm qqH}$, can strengthen the discovery potential in this low mass region. 
The discovery in this region can be strengthened, as suggested by D.Zeppenfeld {\it et.al.} in recent years, 
with an integrated luminosity even smaller than 30 fb$^{-1}$. 
 Because the minimal supersymmetric extension 
of the standard model (MSSM) predicts the existence of a scalar Higgs 
boson (h or H) with a mass below or around 130 GeV/$c^2$ and with
standard-model like couplings
the results for the SM Higgs boson can be applied directly  to
estimate the potential for the MSSM Higgs bosons.
WBF process is also ideally suited for the detection of an invisibly decaying Higgs boson. 
\begin{2figures}{hbtp}
\resizebox{\linewidth}{78 mm}{\includegraphics{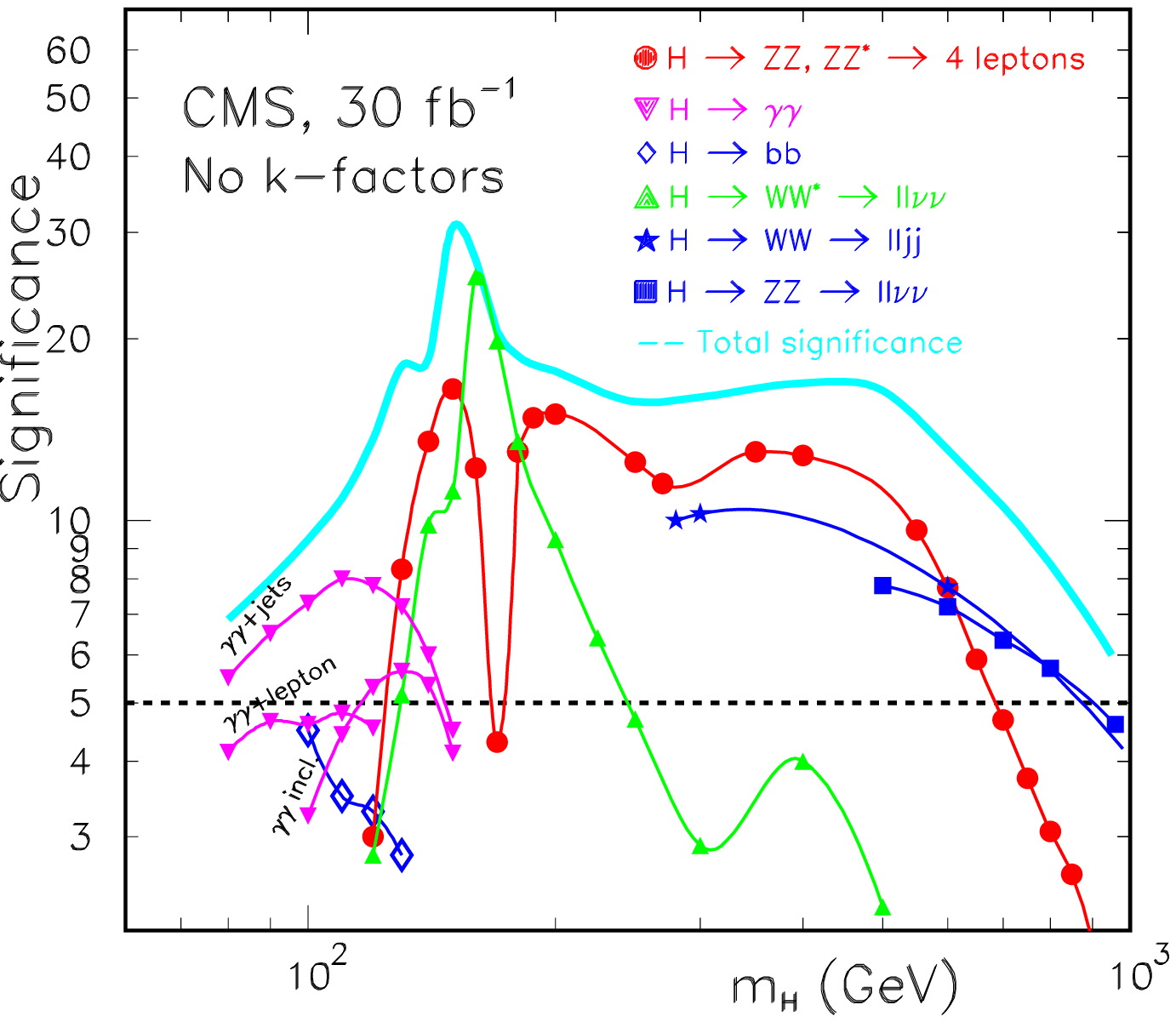}} &
\resizebox{\linewidth}{78 mm}{\includegraphics{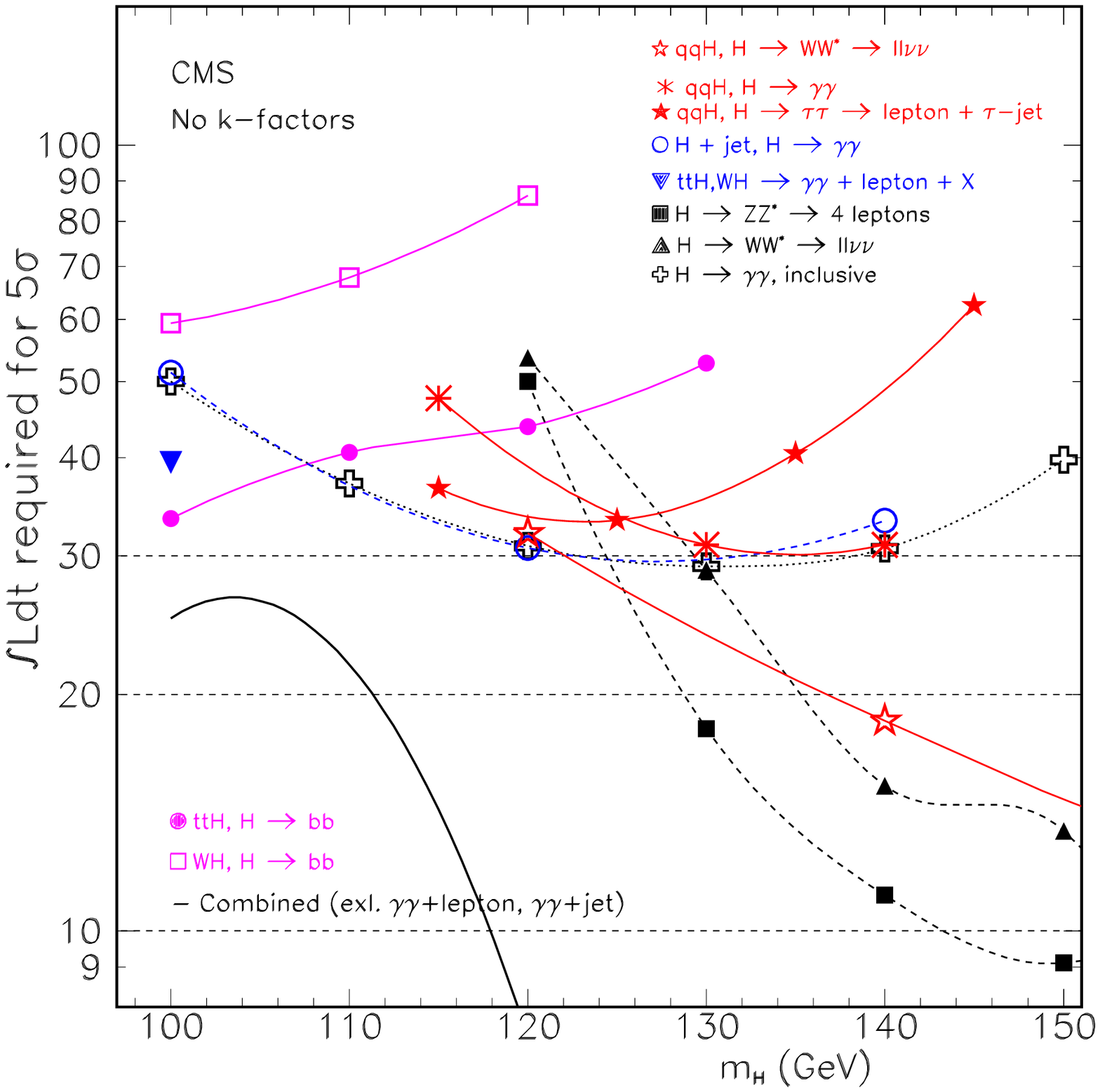}}\\
  \caption{Expected statistical significance ($\rm S / \sqrt{\rm B}$) with 30~fb$^{-1}$ for SM Higgs in CMS with LO cross-sections for all processes.}
  \label{fig:hobs1} &
  \caption{Integrated luminosity required to explore the region $\rm m_{\rm H} \le 150$ GeV/c$^2$ with NLO cross 
sections for the inclusive $\rm H \ra \gamma\gamma$,
 for $\rm H+\rm jet$ with $\rm H \ra \gamma\gamma$ and for $\rm H\ra\rm ZZ^*\ra 4\ell^{\pm}$
 and $\rm H \ra \rm WW^* \ra \ell\ell\nu\nu$.}
  \label{fig:hobs2} 
\end{2figures}
\section{The Higgs boson Search in WBF Process}
The WBF process has distinct signatures which 
provides good handle against potential backgrounds from ${\rm t\bar t}$, 
single W/Z + jets and QCD multi-jet events, though the rate is 
smaller by about an order 
of magnitude compared to the gluon-fusion process. 
In the signal channel the underlying dynamics of simultaneous W or Z
emissions from the incoming quarks and their subsequent fusion to a Higgs boson results
in two energetic jets in the forward and backward regions. The absence of colour
exchange between the scattered quarks and the 
colourless Higgs boson leads to low hadronic activity in the central region, 
when the Higgs boson decays into non-hadronic modes of $ \gamma \gamma$, $\rm WW^\ast \to \rm \ell\nu \rm \ell\nu$, $\tau\tau$, or even invisibly. 
The various QCD background processes are indeed 
 reduced with the requirement of two energetic jets 
({$\rm E^{\rm jet}_{1,2} \ge$ 300 GeV) at large rapidities, with a substantial 
rapidity gap between the jets ($|\Delta \eta| \ge 4$), and that there be very little jet activity 
in the central region: no other jet must be reconstructed with a transverse 
energy in excess of 20 GeV. 
This central-jet-veto efficiency needs to be known to few \%
level and experimentally it is feasible at low-luminosity running condition of LHC ($10^{33}$ cm$^{-2}$ s$^{-1}$). 

The next-to-leading order corrections of WBF is
only $\sim$ 10\% of the leading order cross-section.
 Also the contribution to the $\rm qqH$ final state from gluon-fusion mode, after WBF-specific selection criteria, is 
expected to be only $\sim$ 10\%. These aspects are suitable for the determination of the Higgs boson
couplings when different modes of the Higgs boson production need to be distinguished in various decay final states.


The detection of WBF events demands hermetic calorimetry and the Very Forward 
(VF) Calorimeters of CMS extend up to $|\eta| \le 5$. For about 65\%
of signal events at least one jet lie in VF with the above WBF-specific selection criteria.
Trigger strategies have been extensively studied in CMS for various final state topologies where
the selection thresholds are adjusted to keep reasonable efficiencies for signal channels
while achieving large background rejection factors. The accuracy of the expected signal 
significance depends on the precision of the predicted background rate which sometimes can be measured directly from the data itself using accurately known final states.
\section{Specific Studies in CMS}
\subsection{$\rm qq\rightarrow \rm qqH, H\rightarrow WW^*\rightarrow \ell\nu \ell \nu$}
\label{subsec:lnu}
 In addition to the common general features discussed above, the final 
state for this channel is characterised with two high-$p_T$ leptons and large missing transverse energy, $\etm$~. The potential background from ${\rm t}\bar {\rm t}$ process
has been studied with detailed detector simulation and reconstruction softwares of CMS.
The spin correlation of the two Ws from the Higgs boson decay yields 
an opening angle between the two charged leptons which is smaller, on  
average, than for the background events. The resulting transverse
 mass, reconstructed from the leptons and $\etm$~, has a Jacobian
peak at $m_{\rm H}$ as presented in Fig.~\ref{fig:lnu}. 
This search can cover the Higgs boson mass range from
120 to 170 GeV/c$^2$.
\subsection{$\rm qq\rightarrow \rm qqH, H\rightarrow \gamma\gamma$} 
\label{subsec:gg}
This process has a low signal rate due to the small branching fraction 
($\sim 10^{-3}$) for $\rm H\rightarrow\gamma\gamma$, 
 but much better  signal-to-background ratio 
($S/B \sim 1$ for ${\rm m}_{\rm H}=$ 115 - 140 GeV/c$^2$) than in the gluon fusion mode ($S/B \sim 1/15$). 
This channel provides an interesting complementary final state for an early discovery in the difficult 
low-mass region as illustrated in Fig.~\ref{fig:gmgm}.
\begin{2figures}{hbtp}
  \resizebox{7.2cm}{64 mm}{\includegraphics{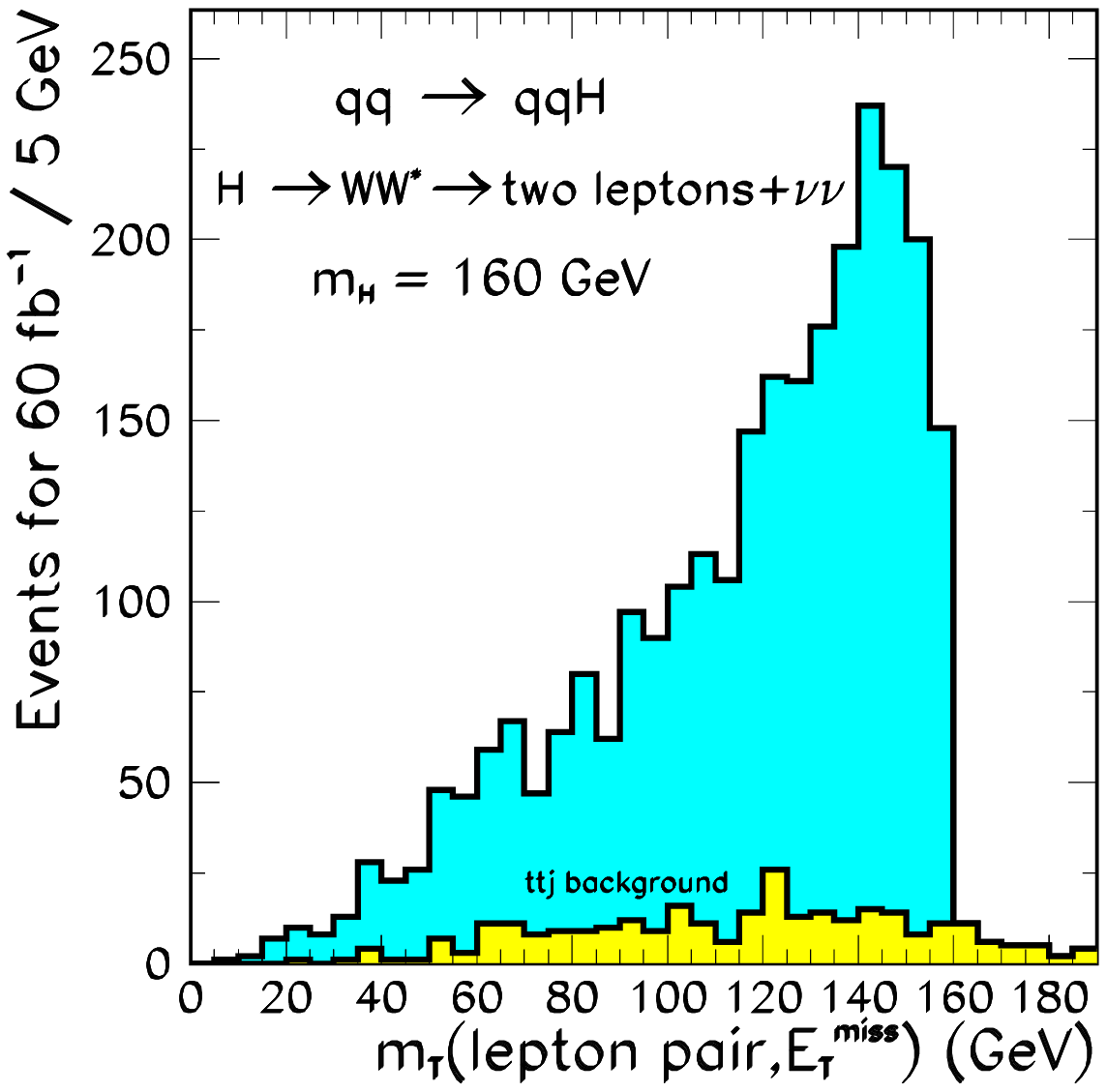}} &
  \resizebox{\linewidth}{64 mm}{\includegraphics{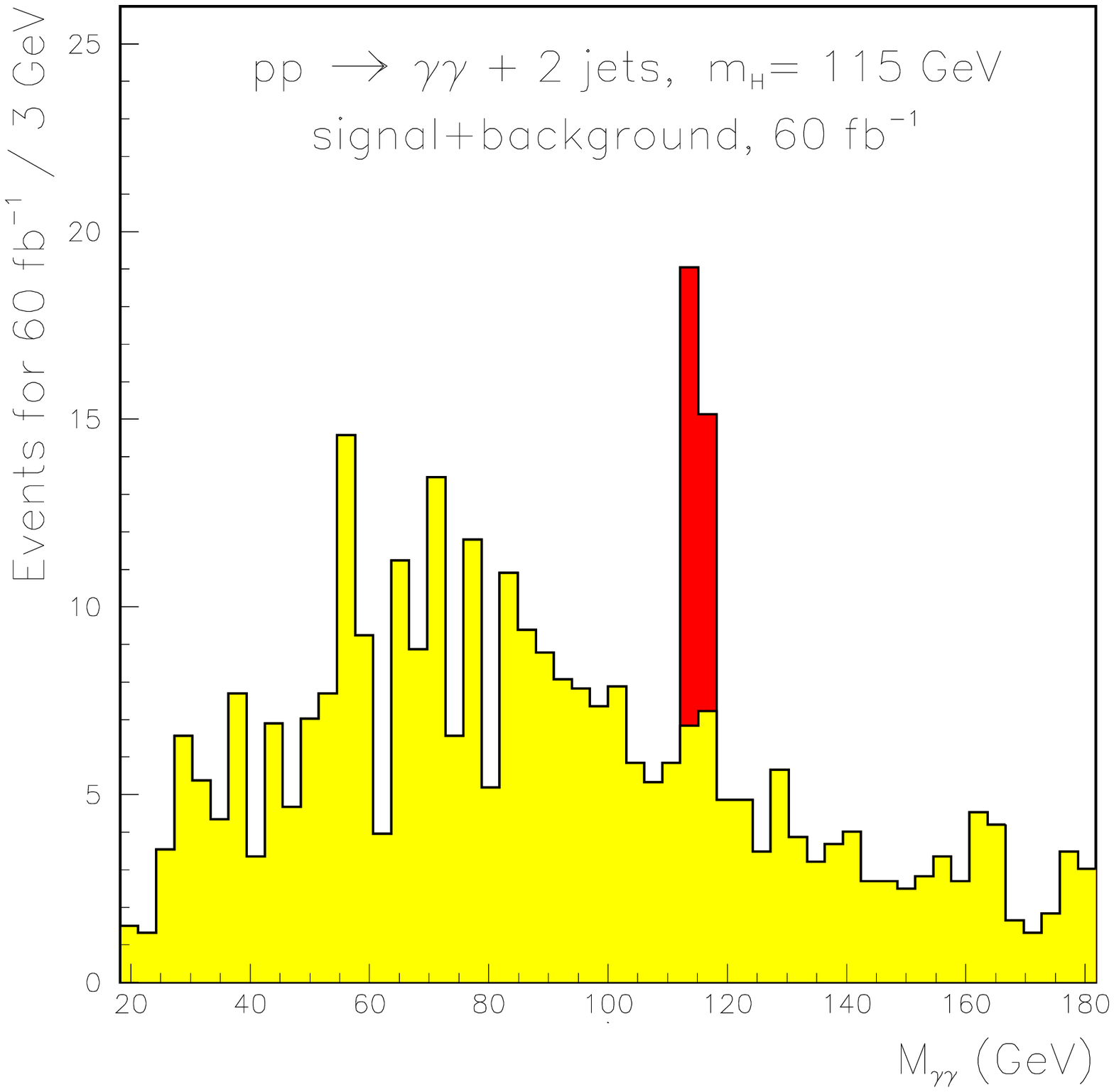}} \\
  \caption{Signal superimposed on background for  
$\rm H\rightarrow \rm WW^*\rightarrow l^+ \nu l^-{\bar\nu}$ for $m_{\rm H}$ = 160 GeV/$c^2$ and 60 fb$^{-1}$.}
\label{fig:lnu} &
\caption{Signal superimposed on background for $\rm H \ra \gamma\gamma$ with 
$m_{\rm H}$ = 120 GeV/$c^2$ and 60 fb$^{-1}$.}
  \label{fig:gmgm}
\end{2figures}
\subsection{$\rm qq\rightarrow \rm qqH, H\rightarrow \tau\tau$} 
\label{subsec:tau}
This channel has been studied thoroughly in CMS. The trigger strategy involves 
$\tau$-identification algorithms based on calorimetric selection at Level 1, and on 
tracking isolations at High Level. 
The Higgs boson mass can be reconstructed from the measured momentum of the 
$\tau$ decay products and the $\etm$ , with the assumption that the latter originates entirely 
from the neutrinos of the $\tau$ decays and is therefore collinear with the parent $\tau$s. In the resulting distribution, shown in 
Fig.~\ref{fig:tau},
the peak is clearly distinguishable 
from the QCD and EW background of $\rm Z+2$ jets which peaks at the Z mass.
\subsection{$\rm qq\rightarrow \rm qqH_{\rm susy}$}
\label{subsec:susy}
One of the neutral Higgs bosons of MSSM, $\rm h$ or $\rm H$, depending on the parameter values, can be searched through its $\tau\tau$ decay mode which alone spans a large region 
of parameter space  even with limited luminosity of
$\sim 30$ fb$^{-1}$. Through WBF production, the channel $\tau\tau \rightarrow \ell + \rm  jet +\etm$  probes the regions 
$m_A \le 120$ GeV/c$^2$ with $\rm H$ and $m_A \ge 150$ GeV/c$^2$ with $\rm h$ as shown in Fig.~\ref{fig:susy}.
\subsection{$\rm qq\rightarrow \rm qqH, H\rightarrow$ Invisible}
In a variety of scenarios beyond the SM, the Higgs boson can decay {\it invisibly}. In this case the transverse momenta of the tagging jets balance the $\etm$ due to the invisible Higgs boson. An upper threshold on the azimuthal angle between the two jets, in addition to the requirement of large $\etm$~, 
reduce the potential background of QCD multijets, heavy-flavours, QCD and EW  W/Z+2 jets types of events.
A sensitivity for the invisible decay branching ratio of the Higgs boson as small as 12\% can be obtained in CMS
 with only 10 fb$^{-1}$ for $m_{\rm H} \lsim 200$ GeV/c$^2$ assuming the vector boson couplings of the {\it invisible} Higgs boson to be the same as in SM.
\begin{2figures}{hbtp}
\resizebox{70mm}{64 mm}{\includegraphics{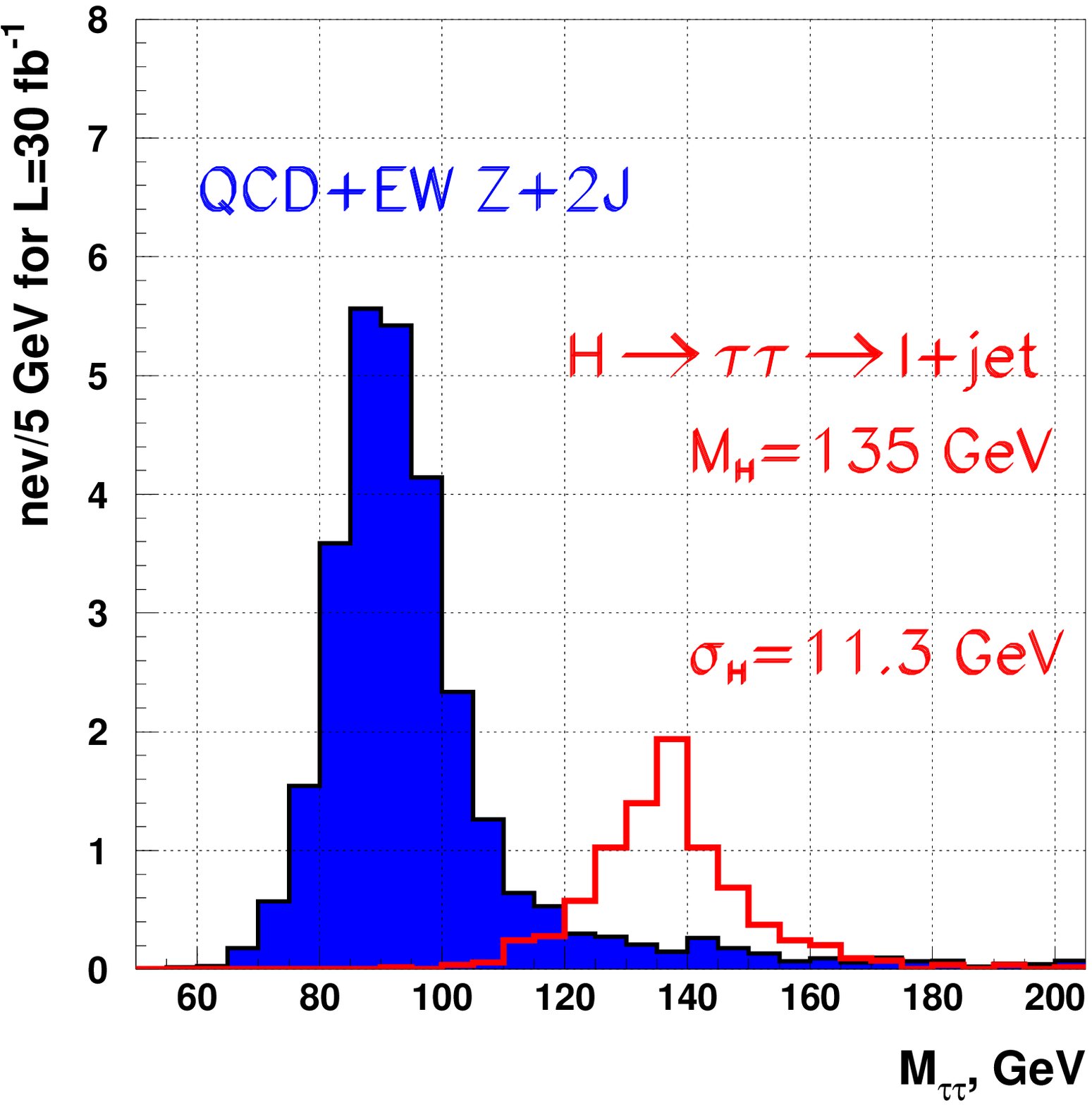}} &
\resizebox{\linewidth}{64 mm}{\includegraphics{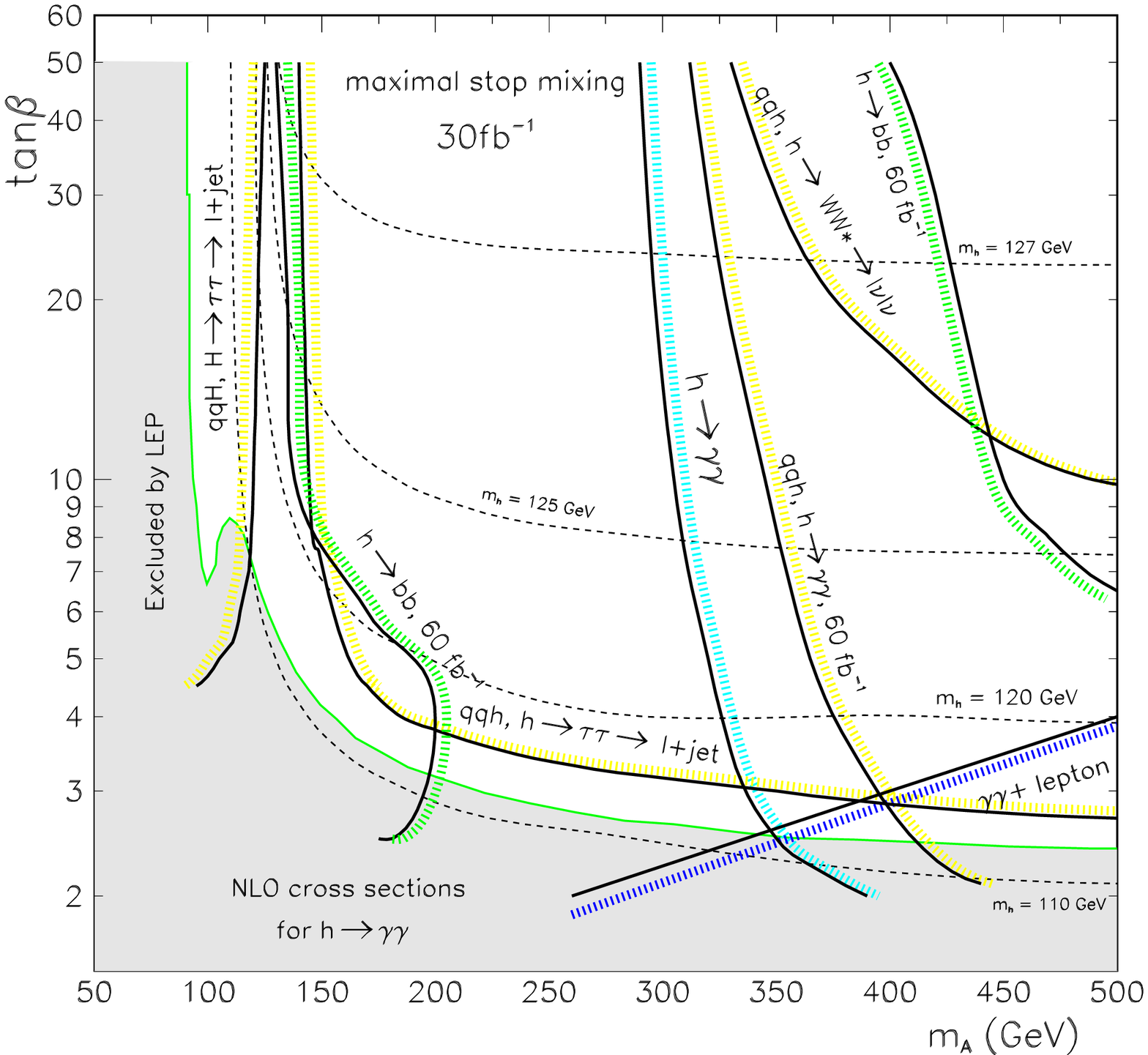}} \\
\caption{Reconstructed mass for SM $\rm H\rightarrow \tau\tau$ with ${\rm m}_{\rm H}=135$ GeV/c$^2$ and backgrounds superposed for 30 fb$^{-1}$. }
\label{fig:tau} & 
\caption{Expected $5\sigma$-discovery range of the neutral scalar (h or H) MSSM Higgs bosons through WBF production in $m_A$-tan$\beta$ parameter space.}
\label{fig:susy}
\end{2figures}
\section{Conclusion}
The search for a low-mass Higgs boson in the SM as well as the neutral Higgs bosons in the MSSM can be enhanced by the weak boson fusion process  with reasonably small integrated luminosity. The distinctive signatures of WBF allows to achieve a statistical significance  similar to that of the gluon-fusion process during the initial low-luminosity phase of LHC.
\section{Acknowledgments} I would like to sincerely thank the organisers for their 
kind hospitality and excellent environment of the conference and 
D.Denegri, A.Nikitenko and  R.Kinnunen for their kind encouragemnet and help. 
\end{document}